\documentclass{JHEP}
\usepackage{epsfig}
\usepackage{amssymb}

\newcommand{\bea}{\begin{eqnarray}}
\newcommand{\eea}{\end{eqnarray}}

\def\d{{\rm d}}

\def\a{{\alpha}}

\def\cpn{${\mathbb{CP}}^{N-1}\,$ }

\preprint{SNUST 060401\\
{\tt hep-th/0604102}}
\title{Emergent AdS$_3$ and BTZ Black Hole\\
from Weakly Interacting Hot 2d CFT~\footnote{This work was supported
in part by the the MOST-KOSEF SRC Program CQUeST (R11-2005-021).} }
\author{Soo-Jong Rey $^a$, Yasuaki Hikida $^b$\\
~~~~~~~~~~~~~~\\
${}^a$ School of Physics and Astronomy \& BK-21 Physics Division\\
Seoul National University, Seoul 151-747 {\rm KOREA}\\
$^b$ Theory Group, KEK, Tsukuba, Ibaraki 305-0801 {\rm JAPAN} \\
~~~~~~~~~~~~~~~~~~~\\
 \email{sjrey@snu.ac.kr, hikida@post.kek.jp} }
\abstract{We investigate emergent holography of weakly coupled
two-dimensional hyperK\"ahler sigma model on $T^* \mathbb{CP}^{N-1}$
at zero and finite temperature. The sigma model is motivated by the
spacetime conformal field theory dual to the near-horizon geometry
of $Q_1$ D1-brane bound to $Q_5$ D5-brane wrapped on $\mathbb{T}^4
\times \mathbb{S}^1$, where $N = Q_1 Q_5$. The sigma model admits
nontrivial instanton for all $N \ge 2$, which serves as a local
probe of emergent holographic spacetime. We define emergent geometry
of the spacetime as that of instanton moduli space via Hitchin's
information metric. At zero temperature, we find that emergent
geometry is AdS$_3$. At finite temperature, time-periodic instanton
is mappable to zero temperature instanton via conformal
transformation. Utilizing the transformation, we show that emergent
geometry is precisely that of the non-extremal, non-rotating BTZ
black hole.}
\begin{document}
\section{Introduction}
In recent work \cite{reyhikida}, exploiting instantons in
four-dimensional ${\cal N}=4$ super Yang-Mills theory, we were able
to extract an emergent holographic dual geometry even at {\sl weak
coupling} $g^2_{\rm YM} \ll 1$ and {\sl small rank} $N \sim {\cal
O}(1)$. Specifically, we examined moduli space geometry of the
Yang-Mills instantons at finite temperature and found that Hitchin's
information metric \cite{hitchin} \footnote{For reviews and further
discussions on information geometry of instantons, see
\cite{infogeo}.}
\bea G^{\rm info}_{AB}(Z) \equiv \int_{\mathbb{R}^4} \Big\langle
{\cal L}_{\rm YM} (\partial_A \log {\cal L}_{\rm YM}) (\partial_B
\log {\cal L}_{\rm YM}) \Big\rangle~, \label{infometric} \eea
of the Yang-Mills instanton density ${\cal L}_{\rm YM}$ exhibits
features identifiable as a Schwarzschild-like black hole whose
geometry asymptotes to five-dimensional anti-de sitter space at
infinity. Given that this is the regime outside Maldacena's AdS/CFT
correspondence \cite{maldacena}, where string worldsheet and quantum
effects are violent, the findings in \cite{reyhikida} is quite
striking and deserves further investigation.

In this work, we continue pursuing the idea of \cite{reyhikida} but
in a lower dimensional context. Specifically, we consider a class of
$(4,4)$ superconformal field theory, described by two-dimensional
nonlinear hyperK\"ahler sigma model whose target space is $T^*
\mathbb{CP}^{N-1}$, the cotangent bundle of $\mathbb{CP}^{N-1}$. For
$N \rightarrow \infty$, the sigma model is known to emerge in the
infrared limit as holographic dual to the near-horizon geometry of
D1-D5 system wrapped on $\mathbb{T}^4 \times \mathbb{S}^1$ with
$N=Q_1Q_5$, where $Q_1, Q_5$ are the numbers of D1- and D5-branes.
This was first suggested in \cite{maldacena} and further studied in
\cite{d1d5-1}-\cite{d1d5-5}~\footnote{For a recent review on the
subject, see \cite{DMWreview}.}. A natural question then arises: as
for the four-dimensional ${\cal N}=4$ super Yang-Mills theory, is
there a holographic dual geometry emergent out of such sigma model
at weak coupling and at small $N$?

In this work, we shall study geometry of moduli space of holomorphic
instantons for the nonlinear sigma model on $T^*\mathbb{CP}^{N-1}$
by exploiting Hitchin's information metric (\ref{infometric}). We
shall find that the geometry is precisely that of the
three-dimensional anti-de Sitter space at zero temperature and, at
finite temperature, of the non-extremal, non-rotating BTZ black
hole! Given that this is the regime where string worldsheet and
quantum loop corrections are expected to be large, the result
suggests a certain rigidity property of the three-dimensional
anti-de Sitter space and the BTZ black hole when they are emerged as
holographic geometries.

This paper is organized as follows. In section 2, we recall relevant
aspects of nonlinear hyperK\"ahler sigma model on target space
$T^*\mathbb{CP}^{N-1}$ from the holographic dual of near-horizon
geometry of the D1-D5 system. In section 3, we focus on period maps
to the two cycles in the base manifold $\mathbb{CP}^{N-1}$, and
propose to study holomorphic instantons of (2,2) supersymmetric
nonlinear sigma model on $\mathbb{CP}^{N-1}$. In section 4, we study
geometry of the instanton moduli space at zero temperature. We show
that Hitchin's information metric is precisely that of the AdS$_3$.
In section 5, we study sigma model calorons, viz. periodic
instantons at finite temperature with trivial holonomy. The calorons
are straightforwardly constructible from the zero-temperature
instantons via appropriate conformal mapping. Utilizing the map, we
show that Hitchin's information metric is precisely that of
non-extremal, non-rotating BTZ black hole. In section 6, we discuss
a potentially subtle issue in identifying topology of the
'worldsheet' of the D1-D5 effective string, which is specific to
AdS$_3$ and BTZ black hole.

\section{Target Space Conformal Field Theory}
In this section, we shall recapitulate the microscopic theory of
D1-D5 system wrapped on $\mathbb{T}^4$ \cite{d1d5-1}-\cite{d1d5-5}.
Here, the D1-branes are "instanton" strings inside the D5-branes. It
is well known that the moduli space of $Q_1$ instantons in $U(Q_5)$
gauge theory on $\mathbb{T}^4$ is the resolved Sym$_{Q_1 Q_5}
(\widetilde{\mathbb{T}}^4)$. In the small instanton size limit, this
description reduces to the gauge theory corresponding to the system
of $Q_1$ D1-branes and $Q_5$ D5-branes. Here, the D1-branes are
wrapped on $\mathbb{S}^1$ and the D5-branes are wrapped on
$\mathbb{T}^4 \times \mathbb{S}^1$. The relevant limit is when
Vol($\mathbb{T}^4) \sim \alpha'^2$, while Vol($\mathbb{S}^1) \gg
\sqrt{\alpha'}$.

The low-energy dynamics of the above D1-D5 brane complex is
described by two-dimensional (4,4) quiver gauge theory of gauge
group U($Q_1) \times$ U($Q_5$). By construction, the theory lives on
$\mathbb{S}^1$. The gauge theory can also have non-zero $\theta$
angle corresponding to the relative U(1) gauge group. The field
contents arising from massless excitations are as follows. From
(1,1) string, there are one 4d ${\cal N}=2$ vector multiplet $(A_a,
Y_i)$ and one hypermultiplet $Y_m$, all in adjoint representation of
$U(Q_1)$. From (5,5) string, there are again one 4d vector multiplet
$(B_a, X_i)$ and one hypermultiplet $X_m$, all in adjoint
representation of $U(Q_5)$. From (1,5) and (5,1) strings, there
arise a hypermultiplet $\chi = (\chi_1, \chi_2)^{\rm T}$
transforming as a doublet of diagonal SU(2)$_R$ subgroup of internal
SO(4)$_I \simeq$ SU(2)$\times$SU(2) and as a bi-fundamental
representation of U($Q_1$)$\times$U($Q_5)$. Thus, under the relative
U(1) gauge group, $(\chi_1, \chi_2)$ carry charges $(+1, -1)$.

The supersymmetric ground state is determined by two sets of
D-flatness conditions for the adjoints of U($Q_1$) and U($Q_5$),
respectively. We also turn on SU(2)$_R$ triplet Fayet-Iliopoulos
terms ${\bf \zeta} \equiv (\zeta, \zeta_c)$ along the U(1) gauge
subgroups. Decompose the D-flatness conditions irreducibly into
traceless and trace parts. The traceless parts combined with the
SU($Q_1)\times$SU($Q_5$) gauge invariance fixes the two adjoint
hypermultiplets $X_m$ and $Y_m$ completely. The trace parts then put
constraints solely on bi-fundamental hypermultiplets $\chi$, now
deformed by the Fayet-Iliopoulos parameters:
\bea && \chi_1 \otimes \chi_1^* - \chi_2^{\rm T} \otimes \chi_2^{\rm
T*} = \zeta
\nonumber \\
&& \hskip1cm \chi_1 \otimes \chi_2^{\rm T} = \zeta_c~.
\label{u1dflat} \eea

It is now known \cite{seibergwitten1, seibergwitten2} that the
equations (\ref{u1dflat}) define the moduli space
$T^*\mathbb{CP}^{Q_1 Q_5 - 1}$, the cotangent bundle of the
hyper-Kahler manifold $\mathbb{CP}^{Q_1 Q_5 - 1}$. More
specifically, in (\ref{u1dflat}), the first equation for $\chi_1$
describes the base $\mathbb{CP}^{Q_1 Q_5 - 1}$ manifold, while the
second equation for $\chi_2$ defines the cotangent space as the
fiber. The Fayet-Ilopoulos parameter ${\bf \zeta}$ is identifiable
with the moduli of the hyper-Kahler metric on $T^*\mathbb{CP}^{Q_1
Q_5 - 1}$. The simplest situation is when $Q_1 Q_5 =2$, yielding the
Eguchi-Hanson space. In the limit ${\bf \zeta}$ is tuned to zero, it
approaches the orbifold $\mathbb{C}^2/\mathbb{Z}_2$. Likewise, for
$N \equiv Q_1 Q_5 > 2$, the moduli space approaches the symmetric
product orbifold:
\bea T^*\mathbb{CP}^N \quad \longrightarrow \quad
\mbox{Sym}_N(\mathbb{C}^2) = {(\mathbb{C}^2)^N \over \mathbb{S}_N}~.
\label{fitozero}\eea
The singularity that appears in this limit corresponds to the cycle
of length $N$ of the permutation group $\mathbb{S}_N$. It is
associated with the chiral primary operator with dimension
$h=\tilde{h} = (N-1)/2$. We shall be primarily interested in the
moduli subspace associated with the resolution of this cycle. Other
singularities in the moduli space corresponding to other cycles are
orthogonal to the above one, and hence are inherent to the symmetric
product orbifold itself.

Thus, the Higgs branch of the quiver gauge theory corresponds to
U(1) gauged linear sigma model on the hyperK\"ahler moduli space
$T^*\mathbb{CP}^{Q_1 Q_5 - 1}$. This theory is expected to flow in
the infrared to $(4,4)$ supersymmetric CFT$_2$ with central charge
$c = \tilde{c} = 6 Q_1 Q_5$. In particular, integrating out the U(1)
vector multiplet that interacts strongly in the infrared, the theory
is reduced to nonlinear sigma model on $T^*\mathbb{CP}^{Q_1 Q_5 -1}$
\cite{HKLR}.

\section{Holomorphic Instantons in CFT$_2$}
As discussed in the previous section, Fayet-Iliopoulos deformation
resolves the singularity on the Higgs branch and blows it up to a
cycle of length N. In this case, the second homology group of the
cotangent bundle $T^*\mathbb{CP}^{N-1}$ is of rank one, generated by
a 2-cycle inside the base $\mathbb{CP}^{N-1}$. Since the CFT$_2$ at
the infrared, the nonlinear sigma model on $T^*\mathbb{CP}^{N-1}$,
lives on $\mathbb{S}^1$ on which both D1- and D5-branes were
wrapped, there now exists instantons that maps $\Sigma_0 =
\mathbb{S}^1 \times \mathbb{R}_t \simeq \mathbb{C}$ to the base
$\mathbb{CP}^{N-1}$ at zero temperature or $\Sigma_\beta =
\mathbb{S}^1 \times \mathbb{S}^1_\beta \simeq \mathbb{T}_2$ to
$\mathbb{CP}^{N-1}$ at finite temperature $T = 2\pi /\beta$. They
are worldsheet instantons of the effective D1-D5 strings
\footnote{Instantons of this sort were considered from the viewpoint
of 6d supergravity in \cite{mikhailov}.}, equivalently, holomorphic
instantons inside the instanton strings. For example, for the
smallest value of $N-1$, viz. $Q_1 Q_5 = 2$, the target manifold is
the Eguchi-Hanson space. The instanton is the well known harmonic
map to the two cycle blown up from $\mathbb{C}^2/\mathbb{Z}_2$.

We are primarily interested in understanding geometry of the moduli
space of these instantons, framed inside the moduli (sub)space of
the Higgs branch. To this end, we shall describe the holomorphic
worldsheet instantons effectively as instantons in two-dimensional
\cpn model. Evidently, such truncation keeps only the (2,2)
supersymmetry manifest, but this does not affect the conclusions we
shall be drawing. For $\mathbb{CP}^{N-1}$, dim $H^{1,1}(\mathbb{R})
= 1$ and the K\"ahler class is specified by a single parameter.
Correspondingly, the sigma model is specified by a choice of the
K\"ahler potential $K(Z_\a, {Z}^*_\a)$, where $Z_\a, {Z}^*_\a$ are
complex chiral superfields (which just renames $\chi_1$'s). In the
K\"ahler potential, part that is globally defined on
$\mathbb{CP}^{N-1}$ belongs to the D-term, and part whose K\"ahler
form $J = d \wedge \overline{d} K$ shifts complex cohomology classes
belongs to the F-term. To construct instanton solutions, we shall
excite bosonic part of the \cpn model, which consists of $N$ complex
scalar fields $Z_\a$ subject to the constraint $|Z_\a|=1$. It is
given by
\bea S = {N \over \lambda^2} \int_{\Sigma} \, \Big[ \vert \partial_m
Z_\a \vert^2 + {1 \over 4} (Z_\a^* \partial_m Z_\a - Z_\a \partial_m
Z_\a^*)^2 \Big]~. \eea
Here, $\lambda$ is a dimensionless coupling constant. This is is a
unique action involving two derivatives and retaining a local U(1)
invariance $Z_\a (x) \rightarrow e^{ i \varepsilon(x)} Z_\a(x)$.
Alternatively, the model can be defined without the constraint
$|Z|=1$ but with manifest U(1) gauge invariance by introducing a
Lagrange multiplier $\mu^2$ and an abelian gauge potential $A_m$,
respectively. This amounts to defining the $\mathbb{CP}^{N-1}$ sigma
model in terms of gauged $(2,2)$ linear sigma model, which
constructs $\mathbb{CP}^{N-1}$ as a quotient of $\mathbb{C}^N$. In
this formulation, bosonic part of the action reads
\bea S = {N \over \lambda^2} \int_\Sigma \, \Big[ |D_m Z_\a|^2 -
\mu^2 (|Z_\a|^2 - 1) \Big]~, \eea
where $D_m Z_\a \equiv (\partial_m + i A_m ) Z_\a$. The equations of
motion are
\bea (D_m^2 + \mu^2) Z_\a = 0, \qquad \vert Z_\a \vert^2 = 1, \qquad
(Z_\a^* D_m Z_\a - Z_\a D_m Z_\a^*) = 0. \eea
Solving the latter two equations, we find
\bea A_m = {i \over 2} (Z^*_\a \partial_m Z_\a - Z_\a \partial_m
Z^*_\a). \eea
The BPS equation is derivable by rewriting the action under the
condition $|Z_\a|^2 = 1$ as
\bea S = {N \over \lambda^2} \int_\Sigma \Big[ \vert (D_m \mp i
\epsilon_{mn} D_n) Z_\a \vert^2 + 2 \pi \, {\cal F} \Big] \ge {2 \pi
N \over \lambda^2} Q ~, \label{bpsform} \eea
where ${\cal F}$ is the instanton charge density and
\bea Q \equiv  \int_\Sigma {\cal F} =  -{i \over 2 \pi} \int_\Sigma
\epsilon_{mn} (D_m Z_\a)^* (D_n Z_\a) = {1 \over 4 \pi} \int_\Sigma
\epsilon_{mn} F_{mn} \in \mathbb{Z} \eea
is the instanton charge. It is integrally quantized U(1) gauge flux.
From (\ref{bpsform}), the BPS equation follows in the form:
\bea D_m  Z_\a = \pm i \epsilon_{mn} D_n Z_\a~.\label{sdeqn} \eea
Thus, in complex coordinates, components of an instanton
configuration are holomorphic sections of a line bundle over
$\Sigma$. It also follows from (\ref{sdeqn}) that the Lagrange
multiplier is set by the instanton charge density
\bea \mu^2 = - {1 \over 2} \epsilon_{mn} F_{mn}. \eea

The BPS instantons that solve (\ref{sdeqn}) are well known
\cite{gross} - \cite{divecchia}. It is well known that the dimension
of the moduli space for charge $k$ instantons in $\mathbb{CP}^{N-1}$
is given by
\bea \mbox{dim}_{\mathbb{C}} {\cal M}^N_k = k N = k \, Q_1 Q_5~.
\label{modulispace} \eea
We shall now construct holomorphic instantons explicitly in a
suitable parametrization suited for our purpose, both at zero and
finite temperature. We shall then study geometry of the instanton
moduli space by computing Hitchin's information metric.

\section{Emergent AdS$_3$ Geometry at Zero Temperature}

Consider first the holomorphic instantons at zero temperature. The
'worldsheet' of the effective D1-D5 string is $\Sigma_0 =
\mathbb{S}^1 \times \mathbb{R}_t$, which is conformally equivalent
to ${\mathbb{C}}$. Introduce complex coordinates on the worldsheet
$z \equiv (x^1 + i x^2)$, $\overline{z} \equiv (x^1 - i x^2)$. The
most general holomorphic instanton solution of (\ref{sdeqn})
topological charge $1$ is given by
\bea Z_\a = {w_\a \over |w|}, \qquad \mbox{where} \qquad w_\a = (z -
z_o) u_\a + \rho_\a, \qquad u^*_\a \rho_\a = 0, \qquad |u|^2=1.
\label{instanton} \eea
Here, moduli parameters of the holomorphic instanton are $z_o =
x^1_o + i x^2_o, \overline{z}_o = x^1_o - i x^2_o$ for the center,
$\rho = \sqrt{|\rho|^2}$ for the size, and $\rho_\a/\rho$ for the
SU(N) orientation, respectively. The $u_\a$ parametrizes the SU(N)
vacuum. With appropriate SU(N) rotation, we can choose
\bea u_\a = \delta_{\a, 1} \qquad \mbox{and} \qquad \rho_\a = \rho
\, \delta_{\a, 2} \eea
while satisfying the conditions in (\ref{instanton}). Thus, there
are one complex modulus parameter $z_0$ and one reail modulus
parameter $\rho$, specifying center and size of the instanton,
respectively. They range over $z_0 \in \mathbb{C}$ and $\rho \in
\mathbb{R}^+$. We are primarily interested in the three-dimensional
subspace in the moduli space ${\cal M}_k^N$ in (\ref{modulispace})
--- this is the subspace orthogonal to the SU(N) orientation.

The Lagrangian density of the holomorphic instanton reads
\bea {\cal F}(z;z_o) = {1 \over 4 \pi} \epsilon^{mn} F_{mn}(z; z_o).
\eea
Substituting the instanton solution (\ref{instanton}), we find that
\bea {\cal F}(z, z_o) = {1 \over \pi} {|\rho|^2 \over (|z-z_o|^2 +
|\rho|^2)^2}. \eea
Motivated by Hitchin's proposal \cite{hitchin}, we propose
holographic geometry in terms of {\sl quantum-averaged} information
metric of the instanton moduli space as
\bea G^{\rm info}_{AB}(z_o) = \int_\Sigma d z d \overline{z} \,
\Big\langle \cal F} ({\partial_A \log {\cal F}) (\partial_B \log
{\cal F})\Big\rangle~. \label{metricdef} \eea
Here, the bracket refers to normalized path integral average over
the fields $(Z_\a, Z^*_\a)$. Semiclassically, saddle-point
configuration dominates (\ref{metricdef}). By elementary
computations, as was done in \cite{cpnmetric}, the information
geometry spans three-dimensional subspace of $\rho$ and $z_o$. On
this space, the information metric (\ref{metricdef}) is given by
\bea (\d s)^2_{\rm instanton} ={R^2  \over \rho^2} \Big[ {\rm d}
\rho^2 + \d z_o \d \overline{z}_o \Big] ~, \label{adsfinal}\eea
where $R^2 = 4/3$. Had we considered an instanton of topological
charge $k$ instead, the metric remains the same as (\ref{adsfinal}),
except that $R^2 \rightarrow R^2 k^2$.

We see that, as probed by the holomorphic instanton, the
three-dimensional Euclidean anti-de Sitter space emerges as a
holographic geometry of two-dimensional (2,2) supersymmetric sigma
model over $\mathbb{CP}^{N-1}$, which spans relevant part of the
(4,4) hypermultiplet nonlinear sigma model over the hyperK\"ahler
moduli space $T^*\mathbb{CP}^{N-1}$. We also see that the emergent
AdS geometry (\ref{adsfinal}) does not depend on the rank $N = Q_1
Q_5$ and the sigma model coupling $\lambda$, the feature shared by
the information geometry of instantons in four-dimensional ${\cal
N}=4$ superconformal Yang-Mills theory.

\section{Emergent BTZ Black Hole Geometry at Finite Temperature}
Consider next the holomorphic instanton at finite temperature $T = 2
\pi/\beta$. They are sigma model calorons. Adopting the reasoning of
\cite{GPY}, we expect that calorons with {\sl trivial} holonomy
would dominate the thermal partition function at semiclassical
level. In Matsubara formulation, the `worldsheet' of the effective
D1-D5 string is $\Sigma_\beta = \mathbb{S}^1 \times
\mathbb{S}^1_\beta$. For reasons we shall return later, we will open
up $\mathbb{S}^1$ to $\mathbb{R}^1$ of the `worldsheet' and consider
$\Sigma_\beta = \mathbb{R} \times \mathbb{S}^1_\beta$. The
`worldsheet' coordinates $y = y_1 + i y_2$ covers $\Sigma_\beta$
with $y_2 \simeq y_2 + \beta$ identified with compact coordinate on
$\mathbb{S}^1_\beta$. With this choice, the most general unit charge
caloron of trivial holonomy is given by \cite{affleck, actor}
\bea Z_\a = {w_\a \over |w|}, \qquad \mbox{where} \qquad w_\a =
u_\a\, e^{{2\pi \over \beta}(y - y_o)} + v_\a, \quad |u|^2 = |v|^2 =
1, \quad \mbox{arg}(u^*_\a v_\a) = 0~. \label{instantonT} \eea
As before, $u_\a$ parametrizes SU(N) vacuum. Now, in contrast to the
instanton at zero temperature, $v_\a$ is not directly interpretable
as the size parameter $\rho$. This is because $u_\a$ and $v_ \a$ are
no longer SU(N) orthogonal --- the condition arg$(u^*_\a v_\a) = 0$
still leaves Re$(u^*_\a v_\a)$ arbitrary --- and because rescaling
of $v_\a$ is equivalent to shifting $y_0$. By appropriate SU(N)
rotation, we choose to parametrize the moduli as
\bea u_\a = \delta_{\a, 1} \qquad \mbox{and} \qquad v_1 = \sqrt{1 -
a^2}, \quad |v_2| = a, \quad v_3 = \cdots v_N = 0 \quad (0 \le a \le
1)~. \label{caloronmoduli} \eea
This choice is compatible with the conditions on caloron's moduli
parameters given in (\ref{instantonT}).

In fact, the caloron with trivial holonomy can be related to the
instanton at zero temperature. This is most conveniently seen by
making the conformal transformation
\bea  \exp \Big({2 \pi \over \beta} y \Big) = {2 \pi \over \beta}\,
z. \eea
With suitable overall rescaling which leaves $Z_\a$ intact, we can
now rewrite the caloron (\ref{instantonT}) in the form of the zero
temperature instanton (\ref{instanton}):
\bea w_\a = (z - z_o) u_\a + \rho_\a  \eea
by judiciously arranging the moduli parameters so that they satisfy
the constraints (\ref{instanton}). However, this does not mean that
the caloron is the same as the instanton at zero temperature.
Rather, it means that the caloron moduli parameters $(y_0, a)$ are
reinterpretable in terms of the zero temperature instanton moduli
parameters:
\bea \rho_\a = {\beta \over 2 \pi} e^{{2 \pi \over \beta} y_o} [v_\a
- (u^* \cdot v) u_\a], \qquad \rho = a {\beta \over 2\pi} \left\vert
e^{{2 \pi \over \beta} y_0} \right\vert, \qquad z_o = - {\beta \over
2 \pi} \sqrt{1 - a^2} e^{{2 \pi \over \beta} y_0}.
\label{relations}\eea
It indicate that the calorons, once conformally mapped, constitute a
proper subset of all possible instantons at zero temperature.

Now, the complex coordinate $z$ covers the entire complex plane
because of the periodicity of Euclidean time, $y_2 \simeq y_2 +
\beta$. Thus, we can utilize the previous computation of holographic
geometry for the zero temperature instanton to extract information
metric of the caloron. In other words, interpreted in terms of the
instanton moduli $(\rho, z_o)$, Hitchin's information metric for the
caloron is the AdS$_3$ space in the Poincar\'e coordinates. But
then, the information metric of the caloron expressed in terms of
the caloron moduli $(a, y_o)$ is obtainable by substituting the
relations (\ref{relations}). This yields
\bea (\d s)^2_{\rm caloron} &=& {R^2 \over \rho^2} \Big[\d \rho^2 +
\d z_o \d \overline{z}_o\Big]
\nonumber \\
&=& R^2 \Big[{\d a^2 \over a^2 (1 - a^2)} + \Big({2 \pi \over \beta}
\Big)^2 \Big( {\d y_1^2 \over a^2} + \Big({1 \over a^2} - 1\Big) \d
y_2^2\, \Big)\Big]~, \eea
where $y_o \rightarrow y_1 + i y_2, \overline{y}_o \rightarrow y_1 -
i y_2$. Redefining the coordinates as $a = 1/r$, we find that
\bea (\d s)^2_{\rm caloron} = R^2 \Big[{ \d r^2 \over r^2 - 1} +
\Big({2 \pi \over \beta} \Big) \Big( r^2 \d y_1^2 + (r^2 - 1) \d
y_2^2 \Big)\Big]. \label{btz} \eea
Here, because $ a \in (0,1]$, the `radial' variable $r$ ranges only
over $r \in [1, \infty)$. We then recognize that the moduli space
metric (\ref{btz}) is precisely the metric of the non-extremal,
non-rotating BTZ black hole, whose horizon is located at $r_+ = 1$.

Notice that the emergent geometry does not depend on the rank $N =
Q_1 Q_5$ of the hypermultiplet sigma model and the coupling
$\lambda$ at all. This is in contrast to the situation of
four-dimensional ${\cal N}=4$ superconformal Yang-Mills theory,
where the emergent geometry at finite temperature certainly deviated
from the AdS$_5$ Schwarzschild black hole. It suggests that there
exists certain rigidity or nonrenormalization property against
string worldsheet and quantum loop corrections.

\section{Topology}
There remains an issue concerning global aspect of the holographic
geometry, . In constructing instanton and caloron of the
hypermultiplet sigma model as holomorphic worldsheet instantons of
the D1-D5 effective string, we have tacitly taken the `worldsheet'
to have topology of $\Sigma_0 \simeq \mathbb{C}$ and $\Sigma_\beta
\simeq \mathbb{R} \times \mathbb{S}^1_\beta$. Thus, the instanton at
zero temperature is a localized lump and the caloron at finite
temperature is a periodic array around the Euclidean time direction.

On the other hand, by construction, the D1-D5 effective string
theory is defined on $\mathbb{S}^1$, not on $\mathbb{R}$. That would
mean that we should have constructed instantons at zero temperature
as a periodic array around $\mathbb{S}^1$ and calorons at finite
temperature as a doubly period array around $\mathbb{S}^1 \times
\mathbb{S}^1_\beta$. Stated differently, calorons with single or
double periodicity are the relevant configurations in so far as the
boundary condition of hypermultiplets on the `worldsheet' is
concerned. If this were the correct identification of the worldsheet
topology, a technical issue pertains since, according to the theorem
of \cite{nounitcharge}, there is no $k=1$ instantons on
$\mathbb{T}_2$ for any value of $N$, a feature shared with
Yang-Mills theories on $\mathbb{T}^4$ \cite{4dcase}. On the other
hand, the instanton exists on $\mathbb{S}^1 \times \mathbb{R}_t$.
Change of the worldsheet topology arose from turning on finite
temperature. Since this should be a smooth process, we think the
$k=1$ instanton ought to exist not only on $\mathbb{S}^1 \times
\mathbb{R}_t$ but also on $\mathbb{T}_2$.

A possible resolution would be the well-known string theoretic
mechanism that the relative U(1) gauge subgroup is coupled to NS-NS
or R-R two-form potential, and they contribute (generically
non-integral) U(1) flux background on the `worldsheet' of the D1-D5
effective string. Combined with the instanton or caloron
configuration, the gauge-invariant net U(1) flux can be arranged to
zero, and evade conditions of the aforementioned no-go theorem. In
such a situation, the topology of the `worldsheet' can be taken
consistent with the D1-D5 effective string theory, viz. $\Sigma_0 =
\mathbb{S}^1\times\mathbb{R}$ or $\Sigma_\beta = \mathbb{S}^1 \times
\mathbb{S}^1_\beta$, where the size of $\mathbb{S}$ is taken large
compared to the string scale. With such topology, the emergent
geometries described by the metrics (\ref{adsfinal}), (\ref{btz})
indeed describe correctly the global AdS$_3$ and BTZ black holes.

An alternative is to take the correct `worldsheet' topology as
$\Sigma_0 = \mathbb{C}$ at zero temperature and as $\Sigma_\beta =
\mathbb{R} \times \mathbb{S}^1_\beta$ at finite temperature. In this
case, we evade the no-go theorem and have smooth $k=1$ instanton
interpolation with the temperature. But then, the emergent geometry
(\ref{btz}) should be interpreted as describing the BTZ black hole
in extreme high temperature limit only.

We intend to report clarification of this point in a separate work.

\section*{Acknowledgement}
SJR is grateful to David J. Gross, Gautam Mandal and Pierre van Baal
for useful discussions. SJR also acknowledges hospitality of Kavli
Institute for Theoretical Physics and Albert Einstein Institute
while this research was in progress.


\begin{thebibliography}{999}

\bibitem{reyhikida} S.-J. Rey and Y. Hikida, {\sl 5d Black Hole
as Emergent Geometry of Weakly Interacting 4d Hot Yang-Mills Gas},
arXiv:hep-th/0507082.

\bibitem{hitchin} N.J. Hitchin, "The geometry and topology of moduli
spaces", Lecture Notes in Mathematics {\bf 1451}, pp. 1-48
(Springer, Heidelberg, 1988).

\bibitem{infogeo} S. Amari, M.K. Murray and J.M. Rice, "Statistics
and differential geometry", Monographs on Statistics and Applied
Probability {\bf 48} (Chapman and Hall, London, 1993); D. Groisser
and M.K. Murray,
Ann. Glob. Ann. Geom. {\bf 15} (1997) 519-537.

\bibitem{maldacena} J. Maldacena,
Adv. Theor. Math. Phys. {\bf 2} (1998) 231-252
[arXiv:hep-th/9711200].

\bibitem{d1d5-1} A. Giveon, D. Kutasov, N. Seiberg,
Adv. Theor. Math. Phys. {\bf 2} (1998)
733-380 [arXiv:hep-th/9806194].

\bibitem{d1d5-2} J. de Boer, H. Ooguri, H. Robins, J. Tannenhauser,
JHEP {\bf 9812} (1998) 026
[arXiv:hep-th/9812046].

\bibitem{d1d5-3} O. Aharony, M. Berkooz, N. Seiberg,
Adv. Theor. Math. Phys. {\bf 2} (1998) 119-153
[arXiv:hep-th/9712117].

\bibitem{d1d5-4} R. Dijkgraaf, {\sl Instanton Strings and
Hyper-K\"ahler Geometry}, Nucl. Phys. B {\bf 543} (1999) 545-571
[arXiv:hep-th/9810210].

\bibitem{d1d5-5} D. Kutasov, N. Seiberg,
JHEP {\bf 9904} (1999) 008
[arXiv:hep-th/9903219].

\bibitem{DMWreview}
  J.~R.~David, G.~Mandal and S.~R.~Wadia,
  Phys.\ Rept.\  {\bf 369}, 549 (2002)
  [arXiv:hep-th/0203048].

\bibitem{seibergwitten1} N. Seiberg and E. Witten,
JHEP {\bf 9904} (1999) 017 [arXiv:hep-th/9903224].

\bibitem{seibergwitten2}
  N.~Seiberg and E.~Witten,
  JHEP {\bf 9909}, 032 (1999)
  [arXiv:hep-th/9908142].

\bibitem{HKLR}
  N.~J.~Hitchin, A.~Karlhede, U.~Lindstrom and M.~Rocek,
  Commun.\ Math.\ Phys.\  {\bf 108}, 535 (1987).

\bibitem{mikhailov}
A. Mikhailov,
Nucl.\ Phys.\ B {\bf 584}, 545 (2000)
  [arXiv:hep-th/9910126].

\bibitem{gross} D.~J.~Gross,
  Nucl.\ Phys.\ B {\bf 132}, 439 (1978).

\bibitem{witten}
  E.~Witten,
  Nucl.\ Phys.\ B {\bf 149}, 285 (1979).

\bibitem{divecchia}
  A.~D'Adda, P.~Di Vecchia and M.~Luscher,
  Nucl.\ Phys.\ B {\bf 152}, 125 (1979).

\bibitem{cpnmetric}
S. Yahikozawa, Phys. Rev. E {\bf 69} (2004) 026122.

\bibitem{GPY}
D.~J.~Gross, R.~D.~Pisarski and L.~G.~Yaffe,
  Rev.\ Mod.\ Phys.\  {\bf 53}, 43 (1981).

\bibitem{affleck}
I.~Affleck,
  Nucl.\ Phys.\ B {\bf 162}, 461 (1980);
  Nucl.\ Phys.\ B {\bf 171}, 420 (1980).

\bibitem{actor} A. Actor,
Fortschr. Phys. {\bf 33} (1985) 333.

\bibitem{nounitcharge} J.L. Richard and A. Rouet, Nucl. Phys. B {\bf 211} (1983) 447;\\
P. van Baal,
Phys. Lett. B {\bf 48} (1999) 26.

\bibitem{4dcase} S. Mukai, Nagoya Math. J. {\bf 81} (1981) 153;\\
P.J. Braam and P. van Baal, Comm. Math. Phys. {\bf 122} (1989)
122;\\
H. Schenk, Comm. Math. Phys. {\bf 116} (1988) 177.

\end{thebibliography}
\end{document}